\documentclass[11pt]{article}
\usepackage{amsmath}
\usepackage{graphicx}
\usepackage{amsfonts}
\usepackage{theorem}

\newtheorem{Lem}{Lemma}[section]
\newtheorem{Def}[Lem]{Definition}
\newtheorem{The}[Lem]{Theorem}
\newtheorem{Prop}[Lem]{Proposition}
\newtheorem{Cor}[Lem]{Corollary}

\newtheorem{Rem}[Lem]{Remark}

\newtheorem{OP}[Lem]{Open problem}

\input{tcilatex}
\begin{document}

\title{Mathematical inequalities for some divergences }
\author{S. Furuichi$^{1}$\thanks{%
E-mail:furuichi@chs.nihon-u.ac.jp} and F.-C. Mitroi$^{2}$\thanks{%
E-mail:fcmitroi@yahoo.com} \\
$^1${\small Department of Computer Science and System Analysis,}\\
{\small College of Humanities and Sciences, Nihon University,}\\
{\small 3-25-40, Sakurajyousui, Setagaya-ku, Tokyo, 156-8550, Japan}\\
$^{2}${\small University of Craiova, Department of Mathematics,}\\
{\small Street A. I. Cuza 13, Craiova, RO-200585, Romania}}
\date{}
\maketitle

\textbf{Abstract.} Divergences often play important roles for study in
information science so that it is indispensable to investigate their
fundamental properties. There is also a mathematical significance of such
results. In this paper, we introduce some parametric extended divergences
combining Jeffreys divergence and Tsallis entropy defined by generalized
logarithmic functions, which lead to new inequalities. In addition, we give
lower bounds for one-parameter extended Fermi-Dirac and Bose-Einstein
divergences. Finally, we establish some inequalities for the Tsallis
entropy, the Tsallis relative entropy and some divergences by the use of the
Young's inequality. \vspace{3mm}

\textbf{Keywords : } Mathematical inequality, Tsallis relative entropy,
Jeffreys divergence, Jensen-Shannon divergence, Fermi-Dirac divergence,
Bose-Einstein divergence and quasilinear divergence \vspace{3mm}

\textbf{2010 Mathematics Subject Classification : } 94A17 and 26D15 \vspace{%
3mm}

\section{Introduction}

For the study of multifractals, in 1988, Tsallis \cite{Tsa} introduced
one-parameter extended entropy of Shannon entropy by 
\begin{equation}
H_{q}(\mathbf{p})\equiv -\sum_{j=1}^{n}p_{j}^{q}\ln
_{q}p_{j}=\sum_{j=1}^{n}p_{j}\ln _{q}\frac{1}{p_{j}},\,\,(q\geq 0,q\neq 1)
\label{Tsallis_entropy}
\end{equation}%
where $\mathbf{p}=\{p_{1},p_{2},\cdots ,p_{n}\}$ is a probability
distribution with $p_{j}>0$ for all $j=1,2,\cdots ,n$ and the $q-$%
logarithmic function for $x>0$ is defined by $\ln _{q}(x)\equiv \frac{%
x^{1-q}-1}{1-q}$ which uniformly converges to the usual logarithmic function 
$\log (x)$ in the limit $q\rightarrow 1$. Therefore Tsallis entropy
converges to Shannon entropy in the limit $q\rightarrow 1$: 
\begin{equation}
\lim_{q\rightarrow 1}H_{q}(\mathbf{p})=H_{1}(\mathbf{p})\equiv
-\sum_{j=1}^{n}p_{j}\log p_{j}.
\end{equation}%
It is also known that R\'{e}nyi entropy \cite{Ren1961} 
\begin{equation}
R_{q}(\mathbf{p})\equiv \frac{1}{1-q}\log \left(
\sum_{j=1}^{n}p_{j}^{q}\right)  \label{Renyi_entropy}
\end{equation}%
is one -parameter extension of Shannon entropy.

For two probability distributions $\mathbf{p}=\{p_{1},p_{2},\cdots ,p_{n}\}$
and $\mathbf{r}=\{r_{1},r_{2},\cdots ,r_{n}\}$ we have divergences based on
these quantities (\ref{Tsallis_entropy}) and (\ref{Renyi_entropy}). We
denote by 
\begin{equation}
D_{q}(\mathbf{p||r})\equiv \sum_{j=1}^{n}p_{j}^{q}(\ln _{q}p_{j}-\ln
_{q}r_{j})=-\sum_{j=1}^{n}p_{j}\ln _{q}\frac{r_{j}}{p_{j}}
\end{equation}%
Tsallis relative entropy. Tsallis relative entropy converges to the usual
relative entropy (divergence, Kullback-Leibler information) in the limit $%
q\rightarrow 1$: 
\begin{equation}
\lim_{q\rightarrow 1}D_{q}(\mathbf{p||r})=D_{1}(\mathbf{p||r})\equiv
\sum_{j=1}^{n}p_{j}(\log p_{j}-\log r_{j}).
\end{equation}%
We also denote by $R_{q}(\mathbf{p||r})$ the R\'{e}nyi relative entropy \cite%
{Ren1961} defined by 
\begin{equation}
R_{q}(\mathbf{p||r})\equiv \frac{1}{q-1}\log \left(
\sum_{j=1}^{n}p_{j}^{q}r_{j}^{1-q}\right) .
\end{equation}%
Obviously $\lim_{q\rightarrow 1}R_{q}(\mathbf{p||r})=D_{1}(\mathbf{p||r})$.

The divergences can be considered to be a generalization of entropies in the
sense that Shannon entropy can be reproduced by the divergence $\log n-D_{1}(%
\mathbf{p||u})$ for the uniform distribution $\mathbf{u}=\{1/n,1/n,\cdots
,1/n\}$. Therefore the study of divergences it is important for the
developments of information science. In this paper, we study several
mathematical inequalities related to some generalized divergences.

\section{Two parameter entropies and divergences}

In this section and throughout the rest of the paper we consider $\mathbf{p}%
=\{p_{1},p_{2},\cdots ,p_{n}\}$ and $\mathbf{r}=\{r_{1},r_{2},\cdots
,r_{n}\} $ with $p_{j}>0,r_{j}>0$ for all $j=1,2,\cdots ,n$ to be
probability distributions.

We start from the Tsallis quasilinear entropies and Tsallis quasilinear
divergences as they were defined in \cite{FMM2011}.

\begin{Def}[\textbf{\protect\cite{FMM2011}}]
For a continuous and strictly monotonic function $\psi $ on $(0,\infty )$
and $r\geq 0$ with $r\neq 1$ (the nonextensivity parameter), Tsallis
quasilinear entropy ($r$-quasilinear entropy) is defined by 
\begin{equation}
I_{r}^{\psi }(\mathbf{p})\equiv \ln _{r}\psi ^{-1}\left(
\sum_{j=1}^{n}p_{j}\psi \left( \frac{1}{p_{j}}\right) \right) .
\end{equation}
\end{Def}

In this context, as a particular case of Tsallis quasilinear entropy we have
Sharma-Mittal information measure (\cite{Mas2005},\cite{SM1975},\cite{SM1977}%
), that is for $\psi (x)=x^{1-q}$ we have: 
\begin{equation*}
I_{r}^{x^{1-q}}(\mathbf{p})=\ln _{r}\left( \sum_{j=1}^{n}p_{j}^{q}\right) ^{%
\frac{1}{1-q}}=\frac{\left( \sum_{j=1}^{n}p_{j}^{q}\right) ^{\frac{1-r}{1-q}%
}-1}{1-r}=H_{r,q}^{S-M}(\mathbf{p}).
\end{equation*}%
We find that $H_{q,q}^{S-M}(\mathbf{p})=H_{q}(\mathbf{p})$. Sharma-Mittal
entropy is also seen in the literature as a two-parameter extension of R\'{e}%
nyi entropy \cite[Section 5]{ST2007}. This also gives rise to another case
of interest 
\begin{equation}
I_{\frac{2q-1}{q}}^{x^{1-q}}(\mathbf{p})=\ln _{\frac{2q-1}{q}}\left(
\sum_{j=1}^{n}p_{j}^{q}\right) ^{\frac{1}{1-q}}=\frac{q}{1-q}\left[ \left(
\sum_{j=1}^{n}p_{j}^{q}\right) ^{\frac{1}{q}}-1\right] ,
\end{equation}%
which coincides with Arimoto's entropy for $q=1/\beta ,$ cf. \cite{Ari1971},
and with $R$-norm information measure, for $R=q,$ cf. \cite{BL1980}.

\begin{Def}[\textbf{\protect\cite{FMM2011}}]
\label{def_TQ}\textbf{\ }For a continuous and strictly monotonic function $%
\psi $ on $(0,\infty )$, the Tsallis quasilinear relative entropy is defined
by 
\begin{equation}
D_{r}^{\psi }(\mathbf{p||r})\equiv -\ln _{r}\psi ^{-1}\left(
\sum_{j=1}^{n}p_{j}\psi \left( \frac{r_{j}}{p_{j}}\right) \right) .
\end{equation}
\end{Def}

Sharma-Mittal divergence (\cite{ABS2007},\cite{Mas2006}) becomes now a
particular case of Tsallis quasilinear divergence: 
\begin{eqnarray*}
&&D_{r}^{x^{1-q}}(\mathbf{p||r})=-\ln _{r}\left( \sum_{j=1}^{n}p_{j}\left( 
\frac{r_{j}}{p_{j}}\right) ^{1-q}\right) ^{\frac{1}{1-q}}=-\ln _{r}\left(
\sum_{j=1}^{n}p_{j}^{q}r_{j}^{1-q}\right) ^{\frac{1}{1-q}} \\
&=&\frac{-\left\{ \left[ \left( \sum_{j=1}^{n}p_{j}^{q}r_{j}^{1-q}\right) ^{%
\frac{1}{1-q}}\right] ^{1-r}-1\right\} }{1-r}=\frac{1-\left(
\sum_{j=1}^{n}p_{j}^{q}r_{j}^{1-q}\right) ^{\frac{1-r}{1-q}}}{1-r}%
=D_{r,q}^{S-M}(\mathbf{p||r}).
\end{eqnarray*}%
By analogy to the entropy computation, we find the following Arimoto type
divergence:%
\begin{equation}
D_{\frac{2q-1}{q}}^{x^{1-q}}(\mathbf{p||r})=\frac{q}{1-q}\left[ 1-\left(
\sum_{j=1}^{n}p_{j}^{q}r_{j}^{1-q}\right) ^{\frac{^{1}}{q}}\right] .
\end{equation}

\begin{Rem}
In limit $r\rightarrow 1$ we have $I_{r}^{x^{1-q}}(\mathbf{p})\rightarrow
I_{1}^{x^{1-q}}(\mathbf{p})=R_{q}(\mathbf{p})$ and $D_{r}^{x^{1-q}}(\mathbf{%
p||r})\rightarrow D_{1}^{x^{1-q}}(\mathbf{p||r})=R_{q}(\mathbf{p||r}).$ It
is known that for $q\neq r$ the Sharma-Mittal divergence fails to conform to
Shore-Johnson theorem \cite{SR81,SR82,SR83}, that is Sharma-Mittal
divergence cannot be written as a $f-$divergence%
\begin{equation*}
D_{r,q}^{S-M}(\mathbf{p||r})=\sum_{j=1}^{n}p_{j}f\left( \frac{r_{j}}{p_{j}}%
\right) ,
\end{equation*}%
for some function $f.$ The previous limits give us a very intuitive way to
conclude that R\'{e}nyi divergence has a similar failure \cite{ABS2007}.
Also this enables us to say that the two-parameter extended relative entropy
discussed in \cite[Section 6]{SF2010} cannot be seen as a particular case of
Sharma-Mittal divergence.
\end{Rem}

\begin{Rem}
\label{rem_SMTR}For $x>0$ and $r\geq 0$ with $r\neq 1$, we define the $r$%
-exponential function as the inverse function of the $r$-logarithmic
function by $\exp _{r}(x)\equiv \left\{ 1+(1-r)x\right\} ^{1/(1-r)}$, if $%
1+(1-r)x>0$, otherwise it is undefined. Here is another connection among
Sharma-Mittal entropy, Tsallis entropy and R\'{e}nyi entropy \cite{FMM2011};%
\cite[(B.8)]{SW2008}: 
\begin{equation*}
\exp _{r}H_{r,q}^{S-M}(\mathbf{p})=\exp _{q}H_{q}(\mathbf{p})=\exp R_{q}(%
\mathbf{p}).
\end{equation*}%
As for a connection among their divergences, we get 
\begin{eqnarray*}
\exp _{2-r}\left( D_{r,q}^{S-M}(\mathbf{p||r})\right) &=&\left\{
1+(r-1)D_{r,q}^{S-M}(\mathbf{p||r})\right\} ^{1/(r-1)}=\left\{ 1+(q-1)D_{q}(%
\mathbf{p||r})\right\} ^{1/(q-1)} \\
&=&\exp _{2-q}\left( D_{q}(\mathbf{p||r})\right) =\exp R_{q}(\mathbf{p||r}).
\end{eqnarray*}
\end{Rem}

\begin{Rem}
The weighted quasilinear mean for some continuous and strictly monotonic
function $\psi :I\rightarrow \mathbb{R}$ is defined by 
\begin{equation}
M_{\left[ \psi \right] }(x_{1},x_{2},\cdots ,x_{n})\equiv \psi ^{-1}\left(
\sum_{j=1}^{n}p_{j}\psi (x_{j})\right) ,
\end{equation}%
where $\sum_{j=1}^{n}p_{j}=1$, $p_{j}>0$, $x_{j}\in I$ for $j=1,2,\cdots ,n$%
. \ It is known that $M_{\left[ \psi \right] }(x_{1},x_{2},\cdots ,x_{n})=M_{%
\left[ \varphi \right] }(x_{1},x_{2},\cdots ,x_{n})$ if and only if \ $\psi
\ $and $\varphi \ $are affine maps of each other, i.e. there exist constants 
$a,$ $b$ such that\ $\psi =a\varphi +b\ $ (cf. \cite[page 141]{AD1975}, cf.
also \cite{Duk2010}). We conclude that $M_{\left[ x^{1-q}\right]
}(x_{1},x_{2},\cdots ,x_{n})=M_{\left[ \ln _{q}\right] }(x_{1},x_{2},\cdots
,x_{n}),$ a fact that yields $I_{r}^{x^{1-q}}(\mathbf{p})=I_{r}^{\ln _{q}}(%
\mathbf{p})=H_{r,q}^{S-M}(\mathbf{p})$ and $D_{r}^{x^{1-q}}(\mathbf{p||r}%
)=D_{r}^{\ln _{q}}(\mathbf{p||r})=D_{r,q}^{S-M}(\mathbf{p||r}).$
\end{Rem}

\section{Jeffreys and Jensen-Shannon type divergences}

\subsection{Tsallis type divergences}

We firstly review the definitions of two famous divergences.

\begin{Def}[\protect\cite{Dra2000},\protect\cite{Jef1946}]
\label{def_J_JS}The Jeffreys divergence is defined by%
\begin{equation}
J_{1}(\mathbf{p||r})\equiv D_{1}(\mathbf{p||r})+D_{1}(\mathbf{r||p})
\end{equation}%
and the Jensen-Shannon divergence is defined as%
\begin{equation}
JS_{1}(\mathbf{p||r})\equiv \frac{1}{2}D_{1}\left( \mathbf{p||}\frac{\mathbf{%
p+r}}{2}\right) +\frac{1}{2}D_{1}\left( \mathbf{r||}\frac{\mathbf{p+r}}{2}%
\right) .
\end{equation}
\end{Def}

Analogously we may define the following divergences.

\begin{Def}
The Jeffreys-Tsallis divergence is 
\begin{equation}
J_{r}(\mathbf{p||r})\equiv D_{r}(\mathbf{p||r})+D_{r}(\mathbf{r||p})
\end{equation}%
and the Jensen-Shannon-Tsallis divergence is 
\begin{equation}
JS_{r}(\mathbf{p||r})\equiv \frac{1}{2}D_{r}\left( \mathbf{p||}\frac{\mathbf{%
p+r}}{2}\right) +\frac{1}{2}D_{r}\left( \mathbf{r||}\frac{\mathbf{p+r}}{2}%
\right) .
\end{equation}
\end{Def}

We find that $J_{r}(\mathbf{p||r})=J_{r}(\mathbf{r||p})$ and $JS_{r}(\mathbf{%
p||r})=JS_{r}(\mathbf{r||p})$. That is, these divergences are symmetric in
the above sense.

To show one of main results in this paper, we need the following lemma that
has interest on its own.

\begin{Lem}
\label{lemma_1}The function 
\begin{equation*}
f\left( x\right) =-\ln _{r}\frac{1+\exp _{q}\left( -x\right) }{2}
\end{equation*}%
is concave\ for $0\leq r\leq q$.
\end{Lem}

\textit{Proof}: The proof is a straightforward computation. The second
derivative is given by 
\begin{eqnarray*}
f^{\prime \prime }(x) &=&-2^{r-1}\left\{ 1+(q-1)x\right\} ^{\frac{2q-1}{1-q}%
}\left( 1+\left\{ 1+(q-1)x\right\} ^{\frac{1}{1-q}}\right) ^{-r-1} \\
&&\times \left( q+(q-r)\left\{ 1+(q-1)x\right\} ^{\frac{1}{1-q}}\right) .
\end{eqnarray*}%
Therefore if $q\geq r$, then the function $f(x)$ is concave.

\hfill \hbox{\rule{6pt}{6pt}}

We wish to note here that the above result yields the fact that under the
same conditions the function $-\ln _{r}\frac{1+\exp _{q}\left( x\right) }{2}$
is also concave, as the composition of a concave function with an affine one.

\begin{Lem}
\label{div}Tsallis divergence satisfies%
\begin{equation*}
D_{r}\left( \mathbf{p||}\frac{\mathbf{p+r}}{2}\right) \leq \frac{1}{2}D_{%
\frac{1+r}{2}}(\mathbf{p||r}).
\end{equation*}
\end{Lem}

\textit{Proof}: From the famous inequality between the arithmetic and
geometric means, we have%
\begin{equation*}
\frac{p_{j}+r_{j}}{2}\geq \sqrt{p_{j}r_{j}}
\end{equation*}%
for all $j=1,2,\cdots ,n$. This implies that%
\begin{eqnarray*}
D_{r}\left( \mathbf{p||\frac{p+r}{2}}\right)  &=&-\sum_{j=1}^{n}p_{j}\ln
_{r}\left( \frac{\frac{p_{j}+r_{j}}{2}}{p_{j}}\right) \leq
-\sum_{j=1}^{n}p_{j}\ln _{r}\left( \frac{\sqrt{p_{j}r_{j}}}{p_{j}}\right)
=-\sum_{j=1}^{n}p_{j}\ln _{r}\left( \sqrt{\frac{r_{j}}{p_{j}}}\right)  \\
&=&-\sum_{j=1}^{n}p_{j}\frac{\left( \sqrt{\frac{r_{j}}{p_{j}}}\right)
^{1-r}-1}{1-r}=-\sum_{j=1}^{n}p_{j}\frac{\left( \frac{r_{j}}{p_{j}}\right)
^{1-\frac{1+r}{2}}-1}{1-r}=-\frac{1}{2}\sum_{j=1}^{n}p_{j}\frac{\left( \frac{%
r_{j}}{p_{j}}\right) ^{1-\frac{1+r}{2}}-1}{1-\frac{1+r}{2}} \\
&=&\frac{1}{2}D_{\frac{1+r}{2}}(\mathbf{p||r}).
\end{eqnarray*}

\hfill \hbox{\rule{6pt}{6pt}}

Hence we derive the following result.

\begin{The}
It holds that%
\begin{equation}
JS_{r}(\mathbf{p||r})\leq \min \left\{ -\ln _{r}\frac{1+\exp _{q}\left( -%
\frac{1}{2}J_{q}(\mathbf{p||r})\right) }{2},\frac{1}{4}J_{\frac{1+r}{2}}(%
\mathbf{p||r})\right\}  \label{js_r}
\end{equation}%
for $0\leq r\leq q$.
\end{The}

\textit{Proof}: According to Lemma \ref{lemma_1},%
\begin{eqnarray}
JS_{r}(\mathbf{p||r}) &\equiv &\frac{1}{2}\left( -\sum_{j=1}^{n}p_{j}\ln _{r}%
\frac{1+\exp _{q}\ln _{q}\left( \frac{r_{j}}{p_{j}}\right) }{2}%
-\sum_{j=1}^{n}r_{j}\ln _{r}\frac{1+\exp _{q}\ln _{q}\left( \frac{p_{j}}{%
r_{j}}\right) }{2}\right)  \notag \\
&\leq &\frac{1}{2}\left( -\ln _{r}\frac{1+\exp _{q}\sum_{j=1}^{n}p_{j}\ln
_{q}\left( \frac{r_{j}}{p_{j}}\right) }{2}-\ln _{r}\frac{1+\exp
_{q}\sum_{j=1}^{n}r_{j}\ln _{q}\left( \frac{p_{j}}{r_{j}}\right) }{2}\right)
\notag \\
&=&\frac{1}{2}\left( -\ln _{r}\frac{1+\exp _{q}\left( -D_{q}(\mathbf{p||r}%
)\right) }{2}-\ln _{r}\frac{1+\exp _{q}\left( -D_{q}(\mathbf{r||p})\right) }{%
2}\right) .
\end{eqnarray}%
Then%
\begin{equation}
JS_{r}(\mathbf{p||r})\leq -\ln _{r}\frac{1+\exp _{q}\left( \frac{-D_{q}(%
\mathbf{p||r})-D_{q}(\mathbf{r||p})}{2}\right) }{2}=-\ln _{r}\frac{1+\exp
_{q}\left( -\frac{1}{2}J_{q}(\mathbf{p||r})\right) }{2}.  \notag
\end{equation}%
We apply Lemma \ref{div} whence it follows%
\begin{equation*}
JS_{r}(\mathbf{p||r})\leq \frac{1}{4}\left( D_{\frac{1+r}{2}}(\mathbf{p||r}%
)+D_{\frac{1+r}{2}}(\mathbf{r||p})\right) .
\end{equation*}%
Thus the proof is completed.\hfill \hbox{\rule{6pt}{6pt}}

\begin{Rem}
For $q=r,$ $r\rightarrow 1$ we have $JS_{r}(\mathbf{p||r})\rightarrow
JS_{1}, $ $J_{r}(\mathbf{p||r})\rightarrow J_{1}$\ and the inequality (\ref%
{js_r}) gives us 
\begin{equation*}
JS_{1}(\mathbf{p||r})\leq \min \left\{ -\log \frac{1+\exp \left( -\frac{1}{2}%
J_{1}(\mathbf{p||r})\right) }{2},\frac{1}{4}J_{1}(\mathbf{p||r})\right\} .
\end{equation*}%
Since 
\begin{equation*}
-\log \frac{1+\exp \left( -x\right) }{2}\leq \frac{x}{2}
\end{equation*}%
we get the main results in \cite{Cro2008}: 
\begin{equation}
JS_{1}(\mathbf{p||r})\leq -\log \frac{1+\exp \left( -\frac{1}{2}J_{1}(%
\mathbf{p||r})\right) }{2}\leq \frac{1}{4}J_{1}(\mathbf{p||r}).
\label{ineq_Crooks}
\end{equation}
\end{Rem}

\subsection{Dual symmetric divergences}

In this subsection, we introduce another type divergences and then we
highlight some inequalities for them.

\begin{Def}
The dual symmetric Jeffreys-Tsallis divergence and the dual symmetric
Jensen-Shannon-Tsallis divergence are defined by 
\begin{equation}
J_{r}^{(ds)}(\mathbf{p||r})\equiv D_{r}(\mathbf{p||r})+D_{2-r}(\mathbf{r||p}%
)\ 
\end{equation}%
respectively 
\begin{equation}
JS_{r}^{(ds)}(\mathbf{p||r})\equiv \frac{1}{2}\left[ D_{r}\left( \mathbf{p||}%
\frac{\mathbf{p+r}}{2}\right) +D_{2-r}\left( \mathbf{r||}\frac{\mathbf{p+r}}{%
2}\right) \right] .
\end{equation}
\end{Def}

As one can see directly from the definition, we find that $J_{r}^{(ds)}(%
\mathbf{p||r})=J_{2-r}^{(ds)}(\mathbf{r||p})$ and $JS_{r}^{(ds)}(\mathbf{p||r%
})=JS_{2-r}^{(ds)}(\mathbf{r||p})$. See \cite{SW2008} and references therein
for \textit{additive duality} $r\leftrightarrow 2-r$ in Tsallis statistics.

Then we get the following upper bound for $JS_{r}^{(ds)}(\mathbf{p||r})$.

\begin{Prop}
For $0\leq r\leq 2$, we have 
\begin{equation}
JS_{r}^{(ds)}(\mathbf{p||r})\leq \frac{1}{4}J_{\frac{1+r}{2}}^{(ds)}(\mathbf{%
p||r}).  \label{js_r_new}
\end{equation}
\end{Prop}

\textit{Proof}: We infer from Lemma \ref{div} that 
\begin{equation*}
D_{2-r}(\mathbf{p||}\frac{\mathbf{p+r}}{2})\leq \frac{1}{2}D_{\frac{3-r}{2}}(%
\mathbf{p||r}).
\end{equation*}%
Consequently 
\begin{equation*}
JS_{r}^{(ds)}(\mathbf{p||r})\leq \frac{1}{2}\left( D_{\frac{1+r}{2}}(\mathbf{%
p||r})+D_{\frac{3-r}{2}}(\mathbf{p||r})\right) =\frac{1}{4}J_{\frac{1+r}{2}%
}^{(ds)}(\mathbf{p||r}).
\end{equation*}%
This completes the proof.

\hfill \hbox{\rule{6pt}{6pt}}

In order to derive further results regarding the dual symmetric divercences,
we need the following lemmas.

\begin{Lem}
\label{ds_lem03}The function $\exp _{q}x$ is monotone increasing in $q$, for 
$x\geq 0$.
\end{Lem}

\textit{Proof}: We have 
\begin{equation*}
\frac{\mathrm{d}\exp _{q}x}{\mathrm{d}q}=\frac{\left\{ 1+(1-q)x\right\} ^{%
\frac{q}{1-q}}h_{q}(x)}{(1-q)^{2}},
\end{equation*}%
where 
\begin{equation*}
h_{q}(x)\equiv (q-1)x+\left\{ 1+(1-q)x\right\} \log \left\{ 1+(1-q)x\right\}
.
\end{equation*}%
Then 
\begin{equation*}
\frac{\mathrm{d}h_{q}(x)}{\mathrm{d}x}=(1-q)\log \left\{ 1+(1-q)x\right\}
\geq 0
\end{equation*}%
for $x\geq 0$ and $q\geq 0$. Therefore $h_{q}(x)\geq h_{q}(0)=0$. Thus we
have $\frac{\mathrm{d}\exp _{q}x}{\mathrm{d}q}\geq 0,$ as asserted.

\hfill \hbox{\rule{6pt}{6pt}}

\begin{Lem}
\label{ds_lem02} For $1<r\leq 2$ and $x>0$, we have 
\begin{equation*}
-\ln _{2-r}x\leq -\ln _{r}x\ 
\end{equation*}%
and 
\begin{equation*}
\exp _{2-r}x\leq \exp _{r}x.
\end{equation*}
\end{Lem}

\textit{Proof}: Since we have $x^{1-r}+x^{r-1}\geq 2$, which implies $%
x^{r-1}-1\geq 1-x^{1-r}$, we have for $1<r\leq 2$, 
\begin{equation*}
\ln _{2-r}x=\frac{x^{r-1}-1}{r-1}\geq \frac{1-x^{1-r}}{r-1}=\ln _{r}x.
\end{equation*}%
The second inequality is a consequence of Lemma \ref{ds_lem03}.\hfill %
\hbox{\rule{6pt}{6pt}}

Our next result reads as follows.

\begin{The}
\label{prop_ds}The following inequality holds 
\begin{equation}
\max \left\{ JS_{r}^{(ds)}(\mathbf{r||p}),JS_{r}^{(ds)}(\mathbf{p||r}%
)\right\} \leq -\ln _{r}\frac{1+\exp _{q}\left( -\frac{1}{2}J_{q}(\mathbf{%
p||r})\right) }{2},  \label{JS_ds}
\end{equation}%
for all $1<r\leq 2$ and $r\leq q.$
\end{The}

\textit{Proof}: By Jensen's inequality, applying Lemma \ref{lemma_1}, we
have 
\begin{eqnarray*}
JS_{r}^{(ds)}(\mathbf{p||r}) &\equiv &\frac{1}{2}\left(
-\sum_{j=1}^{n}p_{j}\ln _{r}\frac{1+\exp _{q}\ln _{q}\left( \frac{r_{j}}{%
p_{j}}\right) }{2}-\sum_{j=1}^{n}r_{j}\ln _{2-r}\frac{1+\exp _{q}\ln
_{q}\left( \frac{p_{j}}{r_{j}}\right) }{2}\right) \\
&\leq &\frac{1}{2}\left( -\ln _{r}\frac{1+\exp _{q}\left( -D_{q}(\mathbf{p||r%
})\right) }{2}-\ln _{2-r}\frac{1+\exp _{q}\left( -D_{q}(\mathbf{r||p}%
)\right) }{2}\right) .
\end{eqnarray*}%
Thus, via Lemma \ref{ds_lem02}, it turns out that%
\begin{eqnarray*}
JS_{r}^{(ds)}(\mathbf{p||r}) &\leq &\frac{1}{2}\left( -\ln _{r}\frac{1+\exp
_{q}\left( -D_{q}(\mathbf{p||r})\right) }{2}-\ln _{r}\frac{1+\exp _{q}\left(
-D_{q}(\mathbf{r||p})\right) }{2}\right) \\
&\leq &-\ln _{r}\frac{1+\exp _{q}\left( -\frac{1}{2}J_{q}(\mathbf{p||r}%
)\right) }{2}.
\end{eqnarray*}%
Further we also have $0\leq 2-r<1$ and the computation is similar for $%
JS_{2-r}^{(ds)}(\mathbf{p||r})$, hence we get (using the additive duality) 
\begin{equation*}
JS_{r}^{(ds)}(\mathbf{r||p})=JS_{2-r}^{(ds)}(\mathbf{p||r})\leq -\ln _{r}%
\frac{1+\exp _{q}\left( -\frac{1}{2}J_{q}(\mathbf{p||r})\right) }{2}.
\end{equation*}

\hfill \hbox{\rule{6pt}{6pt}}

\begin{Rem}
For $q=r,$ $r\rightarrow 1$ we have $JS_{r}^{(ds)}(\mathbf{p||r})\rightarrow
JS_{1}^{(ds)}(\mathbf{p||r})=JS_{1}(\mathbf{p||r}),$ $J_{r}^{(ds)}(\mathbf{%
p||r})\rightarrow J_{1}^{(ds)}(\mathbf{p||r})=J_{1}(\mathbf{p||r})$ and the
inequality (\ref{JS_ds}) yields again the left side inequality of (\ref%
{ineq_Crooks}).
\end{Rem}

\begin{Rem}
The inequality (\ref{JS_ds}) does not hold for $0\leq r\leq q<1$, in
general. We have the following counter-example. We consider the probability
distributions $\mathbf{p}=\{2/5,2/5,1/5\}$ and $\mathbf{r}=\{1/10,1/10,4/5\}$%
. Then for $r=q=0.1$, we have 
\begin{equation*}
-\ln _{r}\frac{1+\exp _{q}\left( -\frac{1}{2}J_{q}(\mathbf{p||r})\right) }{2}%
-JS_{r}^{(ds)}(\mathbf{p||r})\simeq -0.141646.
\end{equation*}
\end{Rem}

\begin{OP}
Prove, disprove or find conditions such that the following inequality holds: 
\begin{equation}
JS_{r}^{(ds)}(\mathbf{p||r})\leq -\ln _{r}\frac{1+\exp _{r}\left( -\frac{1}{2%
}J_{r}^{(ds)}(\mathbf{p||r})\right) }{2},\,\,r\in \lbrack 0,2]\backslash
\{1\}.  \label{conj}
\end{equation}%
We have not yet found any counter-example of (\ref{conj}). One may try to
follow the same argument as in the proof of Theorem \ref{prop_ds}. This
means that one should prove 
\begin{align}
& \frac{1}{2}\left( -\ln _{r}\frac{1+\exp _{r}(-D_{r}(\mathbf{p||r}))}{2}%
-\ln _{2-r}\frac{1+\exp _{2-r}(-D_{2-r}(\mathbf{r||p}))}{2}\right)  \notag \\
& \leq -\ln _{r}\frac{1+\exp _{r}\left( -\frac{1}{2}J_{r}^{(ds)}(\mathbf{p||r%
})\right) }{2}.  \label{conj_3}
\end{align}%
For $0\leq r<1$, we have considered already over 100 particular cases
without finding any counter-example for (\ref{conj_3}). For $1<r\leq 2$ we
have the following counter-example. Assume $r=1.3$, $\mathbf{p}%
=\{0.14,0.01,0.85\}$ and $\mathbf{r}=\{0.07,0.48,0.45\}$. Then the right
hand side in (\ref{conj_3}) minus the left hand side in (\ref{conj_3})
approximately equals $-0.0125861$. Therefore in the case of $1<r\leq 2$ the
proof of (\ref{conj}) (if it holds) couldn't begin with Jensen's inequality
as a first step.
\end{OP}

\subsection{More quasilinear divergences}

We generalize the above definitions.

\begin{Def}
Let the quasilinear Jeffreys-Tsallis divergence be 
\begin{equation*}
J_{r}^{\psi }(\mathbf{p||r})\equiv D_{r}^{\psi }(\mathbf{p||r})+D_{r}^{\psi
}(\mathbf{r||p}),
\end{equation*}%
respectively the quasilinear Jensen-Shannon-Tsallis divergence be 
\begin{equation*}
JS_{r}^{\psi }(\mathbf{p||r})\equiv \frac{1}{2}\left[ D_{r}^{\psi }(\mathbf{%
p||}\frac{\mathbf{p+r}}{2})+D_{r}^{\psi }(\mathbf{r||}\frac{\mathbf{p+r}}{2})%
\right] .
\end{equation*}
\end{Def}

The above quasilinear divergences are symmetric in the sense that we have $%
J_{r}^{\psi }(\mathbf{p||r})=J_{r}^{\psi }(\mathbf{r||p})$ and $JS_{r}^{\psi
}(\mathbf{p||r})=JS_{r}^{\psi }(\mathbf{r||p})$. For $\psi (x)=x^{1-q}\ $we
obtain $J_{r}^{x^{1-q}}(\mathbf{p||r})=J_{r}(\mathbf{p||r})$ and $%
JS_{r}^{x^{1-q}}(\mathbf{p||r})=JS_{r}(\mathbf{p||r}).$

\begin{Prop}
\label{js_fi} Let $\psi $ be a continuous and strictly monotonic function on 
$(0,\infty )$. Suppose that $\psi \left( \frac{1+\psi ^{-1}\left( x\right) }{%
2}\right) $ is concave. Then 
\begin{equation*}
JS_{r}^{\psi }(\mathbf{p||r})\leq -\ln _{r}\frac{1+\exp _{q}\left( -\frac{1}{%
2}J_{q}^{\psi }(\mathbf{p||r})\right) }{2},
\end{equation*}%
for all $0\leq r\leq q$.
\end{Prop}

\textit{Proof}: Since 
\begin{eqnarray*}
&&JS_{r}^{\psi }(\mathbf{p||r})\equiv -\frac{1}{2}\ln _{r}\psi ^{-1}\left(
\sum_{j=1}^{n}p_{j}\psi \left( \frac{1+\psi ^{-1}\left( \psi \left( \frac{%
r_{j}}{p_{j}}\right) \right) }{2}\right) \right) \\
&&-\frac{1}{2}\ln _{r}\psi ^{-1}\left( \sum_{j=1}^{n}r_{j}\psi \left( \frac{%
1+\psi ^{-1}\left( \psi \left( \frac{p_{j}}{r_{j}}\right) \right) }{2}%
\right) \right) ,
\end{eqnarray*}%
by Jensen's inequality, due to the monotonicity and from Lemma \ref{lemma_1}%
, we just compute $\ $%
\begin{eqnarray*}
&&JS_{r}^{\psi }(\mathbf{p||r})\leq \frac{1}{2}\left( -\ln _{r}\frac{1+\psi
^{-1}\left( \sum_{j=1}^{n}p_{j}\psi \left( \frac{r_{j}}{p_{j}}\right)
\right) }{2}-\ln _{r}\frac{1+\psi ^{-1}\left( \sum_{j=1}^{n}r_{j}\psi \left( 
\frac{p_{j}}{r_{j}}\right) \right) }{2}\right) \\
&=&\frac{1}{2}\left( -\ln _{r}\frac{1+\exp _{q}\left( -D_{q}^{\psi }(\mathbf{%
p||r})\right) }{2}-\ln _{r}\frac{1+\exp _{q}\left( -D_{q}^{\psi }(\mathbf{%
r||p})\right) }{2}\right) \\
&\leq &-\ln _{r}\frac{1+\exp _{q}\left( -\frac{1}{2}J_{q}^{\psi }(\mathbf{%
p||r})\right) }{2}.
\end{eqnarray*}

\hfill \hbox{\rule{6pt}{6pt}}

\section{Fermi-Dirac and Bose-Einstein type divergences}

As one-parameter extension of Fermi-Dirac entropy and Bose-Einstein entropy
(see also \cite{Kap1983,TPM2008}), that is of 
\begin{equation*}
I_{1}^{FD}(\mathbf{p})\equiv -\sum_{j=1}^{n}p_{j}\log
p_{j}-\sum_{j=1}^{n}(1-p_{j})\log (1-p_{j})
\end{equation*}%
and 
\begin{equation*}
I_{1}^{BE}(\mathbf{p})\equiv -\sum_{j=1}^{n}p_{j}\log
p_{j}+\sum_{j=1}^{n}(1+p_{j})\log (1+p_{j}),
\end{equation*}%
the Fermi-Dirac-Tsallis entropy was introduced in \cite{TPM2008}. Similarly,
we may define the Bose-Einstein-Tsallis entropy.

\begin{Def}
\label{def_FDT_BET_entropy} The Fermi-Dirac-Tsallis entropy is given by 
\begin{equation}
I_{r}^{FD}(\mathbf{p})\equiv \sum_{j=1}^{n}p_{j}\ln _{r}\frac{1}{p_{j}}%
+\sum_{j=1}^{n}\left( 1-p_{j}\right) \ln _{r}\frac{1}{1-p_{j}}
\end{equation}%
and the Bose-Einstein-Tsallis entropy is defined as 
\begin{equation}
I_{r}^{BE}(\mathbf{p})\equiv \sum_{j=1}^{n}p_{j}\ln _{r}\frac{1}{p_{j}}%
-\sum_{j=1}^{n}\left( 1+p_{j}\right) \ln _{r}\frac{1}{1+p_{j}}.
\end{equation}
\end{Def}

Based on the above extensions, we may introduce Fermi-Dirac-Tsallis
divergence and Bose-Einstein-Tsallis divergence in the following way.

\begin{Def}
Let%
\begin{equation}
D_{r}^{FD}(\mathbf{p||r})\equiv -\sum_{j=1}^{n}p_{j}\ln _{r}\frac{r_{j}}{%
p_{j}}-\sum_{j=1}^{n}\left( 1-p_{j}\right) \ln _{r}\frac{1-r_{j}}{1-p_{j}}
\end{equation}%
and 
\begin{equation}
D_{r}^{BE}(\mathbf{p||r})\equiv -\sum_{j=1}^{n}p_{j}\ln _{r}\frac{r_{j}}{%
p_{j}}+\sum_{j=1}^{n}\left( 1+p_{j}\right) \ln _{r}\frac{1+r_{j}}{1+p_{j}}.
\end{equation}%
Then $D_{r}^{FD}$ is called the Fermi-Dirac-Tsallis divergence and $%
D_{r}^{BE}$ is called the Bose-Einstein-Tsallis divergence.
\end{Def}

\begin{Lem}
\label{lemma_FD}For $0<x,y<1$ we have%
\begin{equation}
-x\ln _{r}\frac{y}{x}-(1-x)\ln _{r}\frac{1-y}{1-x}\geq \frac{4^{r}}{r+1}%
\left[ y^{r+1}\left( 1-x\right) ^{r}+x^{r}\left( 1-y\right)
^{r+1}-x^{r}\left( 1-x\right) ^{r}\right] \geq 0.
\end{equation}
\end{Lem}

\textit{Proof}: Following the idea of \cite[Lemma 11.6.1]{CT2006} we denote 
\begin{equation*}
f\left( x,y\right) \equiv -x\ln _{r}\frac{y}{x}-(1-x)\ln _{r}\frac{1-y}{1-x}-%
\frac{4^{r}}{r+1}\left[ y^{r+1}\left( 1-x\right) ^{r}+x^{r}\left( 1-y\right)
^{r+1}-x^{r}\left( 1-x\right) ^{r}\right] .
\end{equation*}%
We get%
\begin{equation*}
\frac{\mathrm{d}f\left( x,y\right) }{\mathrm{d}y}=\left[ \frac{1}{%
y^{r}\left( 1-y\right) ^{r}}-4^{r}\right] \left[ y^{r}\left( 1-x\right)
^{r}-x^{r}\left( 1-y\right) ^{r}\right] .
\end{equation*}%
We can easily check that $\frac{1}{y\left( 1-y\right) }\geq 4$ under the
assumption $0<y<1$. For $y\leq x$ (which implies $y\left( 1-x\right) \leq
x\left( 1-y\right) $) we establish that the function $f$ is decreasing in
its second variable, hence $f\left( x,y\right) \geq f\left( x,x\right) =0$.
Clearly for the case $y\geq x$ we have similarly $\frac{\mathrm{d}f\left(
x,y\right) }{\mathrm{d}y}\geq 0,$ which leads again $f(x,y)\geq f(x,x)=0$.

Our next step is to take 
\begin{equation*}
g(x,y)\equiv y^{r+1}\left( 1-x\right) ^{r}+x^{r}\left( 1-y\right)
^{r+1}-x^{r}\left( 1-x\right) ^{r}.
\end{equation*}%
For $y\geq x$ we may write that 
\begin{equation*}
\frac{\mathrm{d}g\left( x,y\right) }{\mathrm{d}y}=(r+1)\left[
y^{r}(1-x)^{r}-x^{r}(1-y)^{r}\right] \geq 0.
\end{equation*}
Therefore $g(x,y)\geq g(x,x)=0$. For the case of $y\leq x$, one can show
that $g(x,y)\geq 0$ by the similar way.

\hfill \hbox{\rule{6pt}{6pt}}

\begin{Prop}
\label{sec5_prop01} The Fermi-Dirac-Tsallis divergence satisfies 
\begin{equation}
D_{r}^{FD}(\mathbf{p||r})\geq \frac{4^{r}}{r+1}\sum_{j=1}^{n}\left[
r_{j}^{r+1}\left( 1-p_{j}\right) ^{r}+p_{j}^{r}\left( 1-r_{j}\right)
^{r+1}-p_{j}^{r}\left( 1-p_{j}\right) ^{r}\right] \geq 0.
\end{equation}
\end{Prop}

\textit{Proof}: Via Lemma \ref{lemma_FD}, putting $x=p_{j}$ and $y=r_{j}$,
then taking the sum on both sides, it follows the claimed result.

\hfill \hbox{\rule{6pt}{6pt}}

\begin{Lem}
\label{lemma_BE} For $0<x,y<1$ we have%
\begin{equation}
-x\ln _{r}\frac{y}{x}+(1+x)\ln _{r}\frac{1+y}{1+x}\geq \frac{1}{2^{r}\left(
r+1\right) }\left[ \left( 1+x\right) ^{r}y^{r+1}-\left( 1+y\right)
^{r+1}x^{r}+\left( 1+x\right) ^{r}x^{r}\right] \geq 0.
\end{equation}
\end{Lem}

\textit{Proof}: Consider the function 
\begin{equation*}
f(x,y)\equiv -x\ln _{r}\frac{y}{x}+(1+x)\ln _{r}\frac{1+y}{1+x}-\frac{1}{%
2^{r}\left( r+1\right) }\left[ \left( 1+x\right) ^{r}y^{r+1}-\left(
1+y\right) ^{r+1}x^{r}+\left( 1+x\right) ^{r}x^{r}\right] .
\end{equation*}%
Differentiating $f$ yields 
\begin{equation*}
\frac{\mathrm{d}f\left( x,y\right) }{\mathrm{d}y}=\left[ \left( 1+x\right)
^{r}y^{r}-\left( 1+y\right) ^{r}x^{r}\right] \left[ \frac{1}{y^{r}(1+y)^{r}}-%
\frac{1}{2^{r}}\right] .
\end{equation*}%
Obviously one has $y(1+y)\leq 2$ provided $0<y<1$. For the case $x\geq y$
(i.e. $x(1+y)\geq y(1+x)$), we have $\frac{\mathrm{d}f\left( x,y\right) }{%
\mathrm{d}y}\leq 0$ so that $f(x,y)\geq f(x,x)=0$. One checks that for $%
x\leq y$ we get $\frac{\mathrm{d}f\left( x,y\right) }{\mathrm{d}y}\geq 0,$
so that $f(x,y)\geq f(x,x)=0$.

Next, we put 
\begin{equation*}
g(x,y)\equiv \left( 1+x\right) ^{r}y^{r+1}-\left( 1+y\right)
^{r+1}x^{r}+\left( 1+x\right) ^{r}x^{r}.
\end{equation*}%
For $y\geq x$ 
\begin{equation*}
\frac{\mathrm{d}g\left( x,y\right) }{\mathrm{d}y}=(r+1)\left[
y^{r}(1+x)^{r}-x^{r}(1+y)^{r}\right] \geq 0,
\end{equation*}%
whence $g(x,y)\geq g(x,x)=0$. Further, for the case $y\leq x$, one can prove
similarly that $g(x,y)\geq 0$ holds, which ends the proof.

\hfill \hbox{\rule{6pt}{6pt}}

\begin{Prop}
\label{sec5_prop02} The Bose-Einstein-Tsallis divergence satisfies 
\begin{equation}
D_{r}^{BE}(\mathbf{p||r})\geq \frac{1}{2^{r}\left( r+1\right) }\sum_{j=1}^{n}%
\left[ \left( 1+p_{j}\right) ^{r}r_{j}^{r+1}-\left( 1+r_{j}\right)
^{r+1}p_{j}^{r}+\left( 1+p_{j}\right) ^{r}p_{j}^{r}\right] \geq 0.
\end{equation}
\end{Prop}

\textit{Proof}: According to Lemma \ref{lemma_BE}, putting $x=p_{j}$ and $%
y=r_{j}$, then taking the sum on both sides, it follows the claimed result.

\hfill \hbox{\rule{6pt}{6pt}}

\begin{Rem}
Proposition \ref{sec5_prop01} and Proposition \ref{sec5_prop02} give refined
lower bounds for the Fermi-Dirac-Tsallis divergence and the
Bose-Einstein-Tsallis divergence, respectively. At the same time, they
assure the nonnegativity of $D_{r}^{FD}(\mathbf{p||r})$ and $D_{r}^{BE}(%
\mathbf{p||r})$. Tsus we easily find that the following inequality for the
Tsallis relative entropy holds 
\begin{equation*}
D_{r}(\mathbf{p||r})\geq \max \left\{ \sum_{j=1}^{n}\left( 1-p_{j}\right)
\ln _{r}\frac{1-r_{j}}{1-p_{j}},-\sum_{j=1}^{n}\left( 1+p_{j}\right) \ln _{r}%
\frac{1+r_{j}}{1+p_{j}}\right\} .
\end{equation*}
\end{Rem}

\begin{Cor}
The following inequalities hold 
\begin{equation*}
D_{1}^{FD}(\mathbf{p||r})\geq 2\sum_{j=1}^{n}\left( p_{j}-r_{j}\right) ^{2}
\end{equation*}
and 
\begin{equation*}
D_{1}^{EB}(\mathbf{p||r})\geq \frac{1}{4}\sum_{j=1}^{n}\left(
p_{j}-r_{j}\right) ^{2}.
\end{equation*}
Here $D_{1}^{FD}(\mathbf{p||r})$ is called the Fermi-Dirac divergence,
respectively $D_{1}^{BE}(\mathbf{p||r})$ is called the Bose-Einstein
divergence and their definition corresponds to the limit $r\rightarrow 1$ in
Definition \ref{def_FDT_BET_entropy}.
\end{Cor}

\textit{Proof}: Put $r\rightarrow 1$ in Proposition \ref{sec5_prop01} and
Proposition \ref{sec5_prop02}.

\hfill \hbox{\rule{6pt}{6pt}}

\section{Young's inequality and Tsallis entropies with finite sum}

We establish more inequalities involving Tsallis entropy and Tsallis
relative entropy applying Young's inequality.

\begin{Lem}[Young's inequality]
\label{sec4_lem1} Let $m,~n\geq 0~$ and $p,q\in \mathbb{R}$ such that $\frac{%
1}{p}+\frac{1}{q}=1$. If $p<0$ (then $0<q<1$) or $0<p<1$ (then $q<0$), then
one has $\frac{m^{p}}{p}+\frac{n^{q}}{q}\leq mn.$
\end{Lem}

\begin{Lem}
\label{sec4_lem2}

\begin{itemize}
\item[(i)] Let $p,q\in \mathbb{R}$ satisfying $\frac{1}{1-p}+\frac{1}{1-q}=1$%
. If $p>1$ and $0<q<1$, or if $0<p<1$ and $q>1$, then 
\begin{equation*}
\ln _{p}x+\ln _{q}y\leq xy-1.
\end{equation*}

\item[(ii)] Let $p,q\in \mathbb{R}$ satisfying $\frac{1}{p-1}+\frac{1}{q-1}%
=1 $. If $p<1$ and $1<q<2$, or if $1<p<2$ and $q<1$, then 
\begin{equation*}
\ln _{p}\frac{1}{x}+\ln _{q}\frac{1}{y}\geq -xy+1.
\end{equation*}
\end{itemize}
\end{Lem}

\textit{Proof}:

\begin{itemize}
\item[(i)] Using Lemma \ref{sec4_lem1}, we obtain 
\begin{equation*}
\ln _{p}x+\ln _{q}y=\frac{x^{1-p}-1}{1-p}+\frac{y^{1-q}-1}{1-q}\leq xy-1.
\end{equation*}

\item[(ii)] Lemma \ref{sec4_lem1} leads to 
\begin{equation*}
\ln _{p}\frac{1}{x}+\ln _{q}\frac{1}{y}=\frac{x^{p-1}-1}{1-p}+\frac{y^{q-1}-1%
}{1-q}=-\left( \frac{x^{p-1}}{p-1}+\frac{y^{q-1}}{q-1}\right) +\left( \frac{1%
}{p-1}+\frac{1}{q-1}\right) \geq -xy+1.
\end{equation*}
\end{itemize}

\hfill \hbox{\rule{6pt}{6pt}}

Then we have the following proposition.

\begin{Prop}
\label{q_cross}

\begin{itemize}
\item[(i)] Let $p,q\in \mathbb{R}$ satisfying $\frac{1}{1-p}+\frac{1}{1-q}=1$%
. If $1<p<2$ and $0<q<1$, or if $0<p<1$ and $1<q<2$, then 
\begin{equation}
D_{p}(\mathbf{p||r})+H_{2-q}(\mathbf{p})\geq 1-\sum_{j=1}^{n}p_{j}r_{j}
\label{sec4_prop1_01}
\end{equation}%
and 
\begin{equation}
D_{2-p}(\mathbf{p||r})+H_{q}(\mathbf{p})\leq \sum_{j=1}^{n}\frac{p_{j}}{r_{j}%
}-1.  \label{sec4_prop1_02}
\end{equation}

\item[(ii)] Let $p,q\in \mathbb{R}$ satisfying $\frac{1}{p-1}+\frac{1}{q-1}%
=1 $. If $0<p<1$ and $1<q<2$ or if $1<p<2$ and $0<q<1$, then 
\begin{equation}
D_{p}(\mathbf{p||r})+H_{2-q}(\mathbf{p})\leq \sum_{j=1}^{n}\frac{p_{j}}{r_{j}%
}-1  \label{sec4_prop1_03}
\end{equation}%
and 
\begin{equation}
D_{2-p}(\mathbf{p||r})+H_{q}(\mathbf{p})\geq 1-\sum_{j=1}^{n}p_{j}r_{j}.
\label{sec4_prop1_04}
\end{equation}
\end{itemize}
\end{Prop}

\textit{Proof}:

\begin{itemize}
\item[(i)] In (i) of Lemma \ref{sec4_lem2}, since we have $\ln _{q}y=-\ln
_{2-q}\frac{1}{y}$ for all $y>0$, we get%
\begin{equation*}
\ln _{p}x+\ln _{q}y=\ln _{p}x-\ln _{2-q}\frac{1}{y}\leq xy-1.
\end{equation*}%
Putting $x=\frac{r_{j}}{p_{j}}$ and $y=p_{j}$ and multiplying $-p_{j}$ and
then taking the sum on both sides, it follows 
\begin{equation*}
-\sum_{j=1}^{n}p_{j}\ln _{p}\frac{r_{j}}{p_{j}}+\sum_{j=1}^{n}p_{j}\ln _{2-q}%
\frac{1}{p_{j}}\geq \sum_{j=1}^{n}\left( p_{j}-p_{j}r_{j}\right) ,
\end{equation*}%
which implies the inequality (\ref{sec4_prop1_01}). We also have the
inequality (\ref{sec4_prop1_02}) from 
\begin{equation*}
\ln _{p}x+\ln _{q}y=-\ln _{2-p}\frac{1}{x}+\ln _{q}y\leq xy-1.
\end{equation*}

\item[(ii)] Using (ii) of Lemma \ref{sec4_lem2} we have two inequalities (%
\ref{sec4_prop1_03}) and (\ref{sec4_prop1_04}) by the similar way to the
proof of (i).
\end{itemize}

\hfill \hbox{\rule{6pt}{6pt}}

\begin{Rem}
We have a pair of additive duality $(p,2-q) \leftrightarrow (2-p,q)$ between
(i) and (ii) of Proposition \ref{q_cross}.
\end{Rem}

A cross-entropy type formula \cite{MS1999} of two probability distributions
is the following: 
\begin{equation*}
H(\mathbf{p,r})=D_{1}(\mathbf{p||r})+H_{1}(\mathbf{p}).
\end{equation*}

One may see the left side terms in Proposition \ref{q_cross} as some
generalizations of $H(\mathbf{p,r})$.

\begin{Cor}
\label{cor_cross}The following inequalities holds: 
\begin{equation*}
0 \leq 1-\sum_{j=1}^{n}p_{j}r_{j}\leq H(\mathbf{p,r})\leq \sum_{j=1}^{n}%
\frac{p_{j}}{r_{j}}-1.
\end{equation*}
\end{Cor}

\textit{Proof}: In Proposition \ref{q_cross}, we take $q\rightarrow 1$.

\hfill \hbox{\rule{6pt}{6pt}}

\begin{Cor}
The following inequalities hold:%
\begin{equation*}
0\leq 1-\sum_{j=1}^{n}p_{j}^{2}\leq H_{1}(\mathbf{p})\leq n-1.
\end{equation*}
\end{Cor}

\textit{Proof}: In Corollary \ref{cor_cross}, we take $\mathbf{r=p}$.

\hfill \hbox{\rule{6pt}{6pt}}


\begin{Prop}
Let $p,q\in \mathbb{R}$ satisfying $\frac{1}{p-1}+\frac{1}{q-1}=1$. If $p<1$
and $1<q<2$, or if $1<p<2$ and $q<1$, then 
\begin{equation*}
I_{p}^{FD}(\mathbf{p})+I_{q}^{FD}(\mathbf{p})\geq 3\sum_{j=1}^{n}p_{j}\left(
1-p_{j}\right) .
\end{equation*}
\end{Prop}

\textit{Proof}: From Lemma \ref{sec4_lem2}, (ii), putting $x=y=p_{j}$ and
multiplying $p_{j}$ and then taking the sum on both sides, it follows 
\begin{equation*}
\sum_{j=1}^{n}p_{j}\ln _{p}\frac{1}{p_{j}}+\sum_{j=1}^{n}p_{j}\ln _{q}\frac{1%
}{p_{j}}\geq -\sum_{j=1}^{n}p_{j}^{3}+1.
\end{equation*}%
Putting $x=y=1-p_{j}$ and multiplying $1-p_{j}$ and then taking the sum on
both sides, it follows 
\begin{equation*}
\sum_{j=1}^{n}\left( 1-p_{j}\right) \ln _{p}\frac{1}{1-p_{j}}%
+\sum_{j=1}^{n}\left( 1-p_{j}\right) \ln _{q}\frac{1}{1-p_{j}}\geq
-\sum_{j=1}^{n}\left( 1-p_{j}\right) ^{3}+n-1.
\end{equation*}%
Summing up these two inequalities we get 
\begin{equation*}
I_{p}^{FD}(\mathbf{p})+I_{q}^{FD}(\mathbf{p})\geq
n-\sum_{j=1}^{n}p_{j}^{3}-\sum_{j=1}^{n}\left( 1-p_{j}\right)
^{3}=3\sum_{j=1}^{n}p_{j}\left( 1-p_{j}\right) .
\end{equation*}

\hfill \hbox{\rule{6pt}{6pt}}

We also find that the following interesting inequalities on finite sum hold
true.

\begin{Prop}
For two probability distributions $\mathbf{p}=\{p_{1},p_{2},\cdots ,p_{n}\}$
and $\mathbf{r}=\{r_{1},r_{2},\cdots ,r_{n}\}$, we have the following
relations.

\begin{itemize}
\item[(i)] If $0\leq q <1$, then we have $\sum_{j=1}^n p_j^q r_j^{1-q} \leq
1\leq \sum_{j=1}^n p_j^{2-q} r_j^{q-1}$.

\item[(ii)] If $1<q\leq 2$, then we have $\sum_{j=1}^n p_j^q r_j^{1-q} \geq
1 \geq \sum_{j=1}^n p_j^{2-q} r_j^{q-1}$.
\end{itemize}
\end{Prop}

\textit{Proof}: From the nonnegativity of Tsallis relative entropy $D_q(%
\mathbf{p||r}) \geq 0 $ and $D_{2-q}(\mathbf{p||r}) \geq 0$, we have the
statements.

\hfill \hbox{\rule{6pt}{6pt}}

\section{Concluding remarks}

We close this paper giving further generalized entropy and divergence by the
use of two-parameter extended logarithmic function.

\begin{Def}
For a continuous and strictly monotonic function $\psi $ on $(0,\infty )$
and $r,q\geq 0$ with $r,q\neq 1$, the $(r,q)$-quasilinear entropy is defined
by 
\begin{equation}
I_{r,q}^{\psi }(\mathbf{p})\equiv \ln _{r,q}\psi ^{-1}\left(
\sum_{j=1}^{n}p_{j}\psi \left( \frac{1}{p_{j}}\right) \right) .
\end{equation}
\end{Def}

Here the two-parameter extended logarithmic function \cite{ST2007} is given
by $\ln _{r,q}\left( x\right) =\ln _{q}\exp \ln _{r}\left( x\right) .$
Correspondingly, the inverse function of $\ln _{r,q}$ is denoted by $\exp
_{r,q}.$ For $\psi (x)=\ln _{r,q}\left( x\right) $ we recover the entropy
used in \cite[Section 4]{ST2007}.

For $\psi (x)=x^{1-r}$, we have an extension of Tsallis entropy%
\begin{equation*}
I_{r,q}^{x^{1-r}}(\mathbf{p})=\ln _{q}\exp H_{r}(\mathbf{p})\equiv H_{r,q}(%
\mathbf{p}).
\end{equation*}

For $\psi (x)=x^{1-p}$, we have 
\begin{equation*}
I_{r,q}^{x^{1-p}}(\mathbf{p})=\ln _{q}\exp H_{r,p}^{S-M}(\mathbf{p})\equiv
H_{r,q,p}(\mathbf{p})
\end{equation*}%
that extends Sharma-Mittal entropy to a three-parameter entropy.

\begin{Def}
For a continuous and strictly monotonic function $\psi $ on $(0,\infty )$
and $r,q\geq 0$ with $r,q\neq 1$, the $(r,q)$-quasilinear divergence is
defined by 
\begin{equation}
D_{r,q}^{\psi }(\mathbf{p||r})\equiv -\ln _{r,q}\psi ^{-1}\left(
\sum_{j=1}^{n}p_{j}\psi \left( \frac{r_{j}}{p_{j}}\right) \right) .
\end{equation}
\end{Def}

For $\psi (x)=x^{1-r},$ we get the following extension of Tsallis relative
entropy 
\begin{equation*}
D_{r,q}^{x^{1-r}}(\mathbf{p||r})=\ln _{q}\exp D_{r}(\mathbf{p||r})\equiv
D_{r,q}(\mathbf{p||r}).
\end{equation*}

For $\psi (x)=x^{1-p}$, we have 
\begin{equation*}
D_{r,q}^{x^{1-p} }(\mathbf{p||r}) =\ln _{q}\exp D_{r,p}^{S-M}(\mathbf{p||r}%
)\equiv D_{r,q,p}(\mathbf{p||r})
\end{equation*}%
that extends Sharma-Mittal divergence to a three-parameter divergence.

For a three parametrization extension of the logarithmic function see for
instance \cite{Kan2009} and the references cited therein. With such
extensions the quasilinear entropies can be analogously extended to three
parametric classes too. This is not the purpose of the present paper.

\section*{Acknowledgements}

The author (S.F.) was supported in part by the Japanese Ministry of
Education, Science, Sports and Culture, Grant-in-Aid for Encouragement of
Young Scientists (B), 20740067. The author (F.-C. M.) was supported by
CNCSIS Grant $420/2008.$

\end{document}